\documentclass[adp,fleqn]{w-art}
\usepackage{times}
\usepackage{w-thm}
\usepackage{hyperref}
\usepackage{times}
\usepackage{amsmath, amsthm, amssymb}
\usepackage{graphicx}
\usepackage{color}

\newcommand{\Var}{\mathrm{Var}}

\begin{document}

\title{Trajectories without quantum uncertainties}

\author{Eugene S. Polzik\inst{1}}
\address[\inst{1}]{Niels Bohr Institute, Copenhagen University, Blegdamsvej 17, 2100 Copenhagen, Denmark}
\author{Klemens Hammerer\inst{2}}
\address[\inst{2}]{Institute for Theoretical Physics, Institute for Gravitational Physics (Albert Einstein Institute), Leibniz University Hannover, Callinstra\ss{}e 38, 30167 Hannover, Germany}

\begin{abstract}
  A common knowledge suggests that trajectories of particles in quantum mechanics always have quantum uncertainties. These quantum uncertainties set by the Heisenberg uncertainty principle limit precision of measurements of fields and forces, and ultimately give rise to the standard quantum limit in metrology. With the rapid developments of sensitivity of measurements these limits have been approached in various types of measurements including measurements of fields and acceleration. Here we show that a quantum trajectory of one system measured relatively to the other ``reference system'' with an effective negative mass can be quantum uncertainty--free. The method crucially relies on the generation of an Einstein-Podolsky-Rosen entangled state of two objects, one of which has an effective negative mass.  From a practical perspective these ideas open the way towards force and acceleration measurements at new levels of sensitivity far below the standard quantum limit.
\end{abstract}

\date\today

\maketitle

Quantum uncertainties set by the Heisenberg uncertainty principle determine fundamental limits on measurement precision.  With rapid developments of sensitivity of measurements these limits have been approached in various types of measurements including measurements of fields, time, acceleration, and position. It is commonly accepted that the balance between the amount of information obtained from the measurement and the back action of the measurement can be at best balanced such that they result in the standard quantum limit of the position measurement \cite{Braginsky1975,Braginsky1980a,Caves1980a,Caves1981,Braginsky1995a}. Ideas for circumventing this limit have been put forward based on frequency-dependent squeezing \cite{bondurant1984}, variational measurement \cite{Kimble2001}, the use of Kerr media \cite{Bondurant1986}, dual mechanical resonators \cite{Briant2003,Caniard2007}, the optical spring effect \cite{Chen2010}, stroboscopic measurements \cite{Braginsky1995a}, or two-tone measurements \cite{Braginsky1975,Thorne1978,Braginsky1980a,Clerk2008,Suh2013a}.
However, all those apporaches are limited to measurements of a single quadrature operator of a system and hence are intrinsically limited to measurements of the disturbance whose phase is known in advance. In practice the phase of the signal to be detected is likely to be unknown in advance and therefore another approach is desirable.

A new approach to back action cancellation has been demonstrated in Wasilewski et al. \cite{Wasilewski2010} where a carefully engineered quantum measurement on two entangled spin systems has led to partial cancelation of the quantum noise of measurement for a sensor of magnetic fields. The basis for this approach has been an experimental demonstration of an Einstein-Podolsky-Rosen (EPR) state of two atomic spin oscillators \cite{Julsgaard2001}, one of which has an effective negative mass. 
Such an entangled  EPR state can also be created in a hybrid system involving a nano-mechanical oscillator and an atomic spin ensemble with an effective negative mass \cite{Hammerer2009}. A general theoretical frame for such an approach has been very recently formulated \cite{Tsang2010,Tsang2012}. Following these works, proposals for back action evading measurements employing a Bose-Einstein-condensate~\cite{Zhang2013a} or a two-tone drive~\cite{Woolley2013} to realize an oscillator with an effective negative mass have been put forward, and the all-optical implementation suggested in \cite{Tsang2010,Tsang2012} has been studied further \cite{Wimmer2014}.

In this paper we show how using the entangled state of a positive and a negative mass oscillator one can predict the quantum trajectory of the magnetic or mechanical oscillator measured relatively to a specially chosen origin with, in principle, arbitrarily low uncertainty. Our goal is to develop an intuitive physical picture for this new approach towards metrology below the standard quantum limit, and to illustrate it on the basis of the measurements reported in~\cite{Wasilewski2010}. We base our discussion on three principles. First, we state that a trajectory should be defined with respect to some physical origin. Second, we treat this physical origin as a quantum object. Finally, we allow this object to have an effectively negative mass. Under this condition the trajectory defined with respect to this origin can be known to arbitrary precision at any time. Besides being of fundamental interest, this approach opens the way towards force and acceleration measurements at new levels of sensitivity.

\begin{figure}
\center\includegraphics[width=0.7\columnwidth]{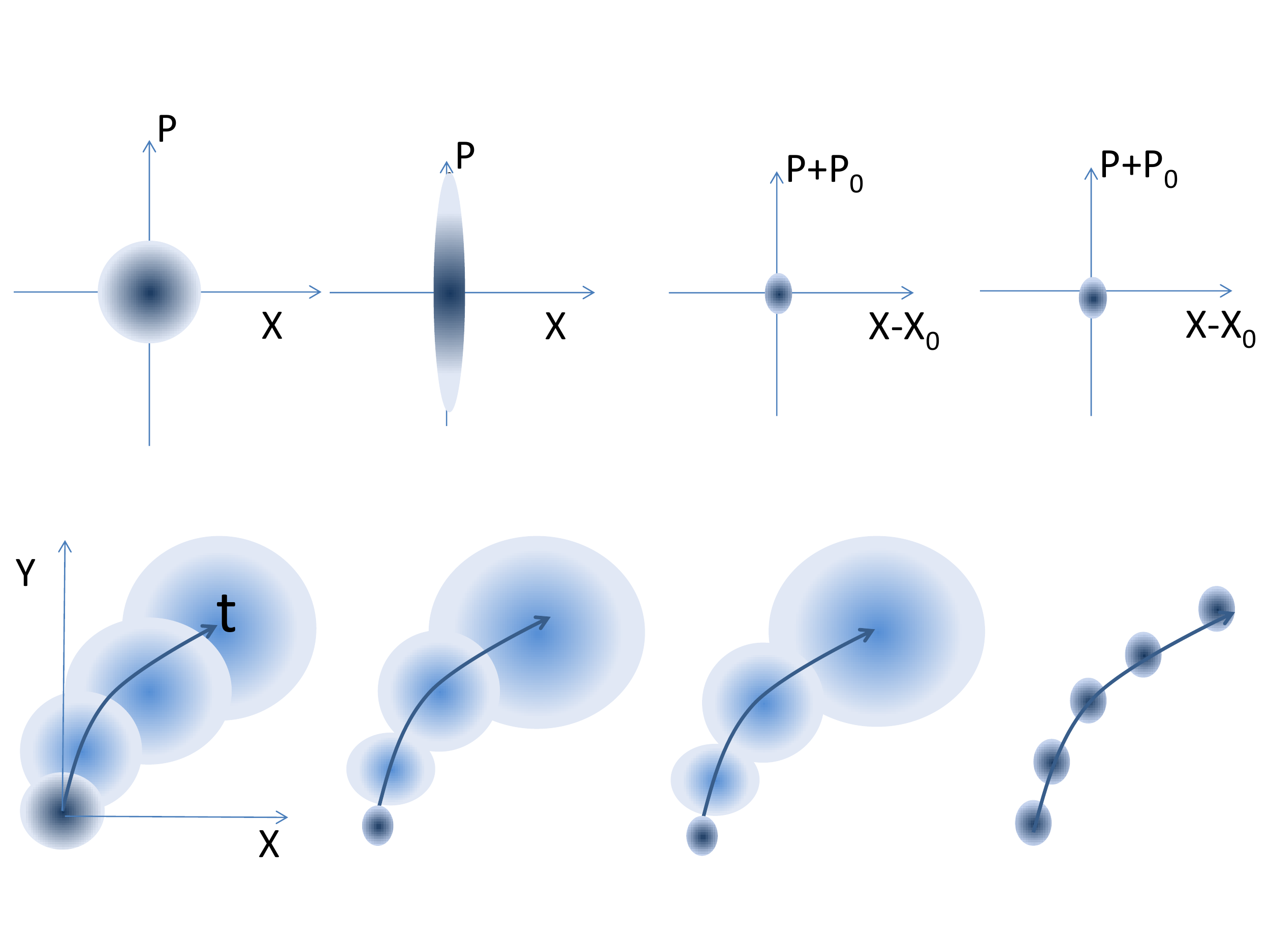}
\caption{\textbf{Trajectories for various quantum states}. Upper row - quantum states of the system, lower row - trajectories. From left to right: coherent state, squeezed state, EPR state with positive mass(frequency), EPR state with negative mass (frequency) }\label{Fig:trajectories}
\end{figure}

According to the rules of quantum mechanics the precision to which we can know the position $X$ of any object is restricted in several ways. First and foremost Heisenberg's principle requires that the uncertainties of position and its conjugate variable, momentum $P$, have to fulfill at any time $\Delta X\Delta P\geq \hbar/2$ as a consequence of the canonical commutator $[X,P]=i\hbar$. This does not of course forbid that the position takes on an arbitrarily sharp value at a particular instant time, provided the momentum is totally blurred at the same moment. Such a squeezed state of motion has been demonstrated for a massive hamrinic oscillator in the form of a trapped ion~\cite{Meekhof1996}. However, as the system evolves -- e.g. $X(t)=X(0)+tP(0)/m$ for a free particle of mass $m$ -- the large uncertainty in momentum will cause a large uncertainty in position at later times. Therefore the trajectory of a freely evolving quantum particle cannot be known to arbitrary precision for arbitrary times, as further explained in Fig.~\ref{Fig:trajectories}.

Fortunately there is a quantum trick which can be played to fight the quantum conspiracy. Note first that the measurement of a position $X$ necessarily is the measurement of a \emph{relative} distance $X-X_0$ between the particle and the origin at $X_0$. Whereas usually this point of reference is taken to be a classical object, we considere a quantum reference frame \cite{Aharonov,Popescu}. This point of reference is considered as a quantum degree of freedom on equal footing with the variable of the system to be measured, such that $[X_0,P_0]=i\hbar$. That is to say, measurement and knowledge of the position of a particle \emph{always} implicitly refers to a relative distance $X-X_0$ between two systems. Assume now that this relative position is measured at time $t=0$. The relative position at future times assuming a free evolution is then given by
$X(t)-X_0(t)=X(0)-X_0(0)+t[P(0)/m-P_0(0)/m_0]$.
 Assume now that the ``quantum origin'' is a very special sort of particle, namely one with a \emph{negative mass}, the physics of which will be detailed below. Moreover, let the mass  be of equal magnitude as the one of the particle to be measured but of opposite sign,
$m_0=-m$. Under this assumption the free evolution of the relative position is \begin{align}\label{eq:evolution}
X(t)-X_0(t)=X(0)-X_0(0)+t[P(0)+P_0(0)]/m.
\end{align}
\emph{Nota bene}, the relative position now couples to the sum of the momenta $P+P_0$. This is an important point because those are commuting quantities, $[X-X_0,P+P_0]=0$, such that they can be known simultaneously with arbitrary precision. The states which possess reduced -- or even vanishing -- variances of relative position and sum of momenta are entangled states of two systems, and coincide with the states considered in the famous Einstein-Podolsky-Rosen argument on the ostensible incompleteness of quantum theory~\cite{Einstein1935}. In order to simplify notations in the following discussion we will assume that the position and momentum variables $X,\,X_0$ and $P,\,P_0$ can be scaled such as to be dimensionless\footnote{In case the two systems are harmonic oscillators of a given frequency $\omega$ the dimensionless canonical coordinates can be naturally introduced by scaling position and momentum to the respective zero point fluctuations $x_{\mathrm{ZPF}}=\sqrt{\hbar/m\omega}$ and $p_{\mathrm{ZPF}}=\sqrt{\hbar m \omega}$}. Therefore, we assume a canonical commutator $[X,P]=[X_0,P_0]=i$ which implies that the variance of each of these canonical variables equals $1/2$ for uncorrelated minimum uncertainty states. In this convention, the degree of entanglement of an Einstein-Podolsky-Rosen state can be characterized by the so-called EPR variance $\Delta_{EPR}=\Var(X-X_0)+\Var(P+P_0)<2$~\cite{Duan2000,Simon2000}. Making an extra crucial assumption of the negative mass of one of the two EPR entangled particles (serving as a quantum reference frame)  allows the relative position to be known to arbitrary precision at all times. Note that generation of an EPR entangled state between a particle and the origin particle with a regular positive effective mass leads to the same uncertainty of the relative position as in the case of a classical origin system (Fig.~\ref{Fig:trajectories}).

We can determine $X-X_0$ in principle but can we actually learn about the relative position without disturbing it? As discussed above, if we measure by some means the relative position $X-X_0$ we will necessarily disturb the conjugate variable, that is $P-P_0$. But this variable does not couple to the relative position at later times in case of the negative mass origin particle, see Eq.~\eqref{eq:evolution}. The back action effect of the  measurement therefore cannot spoil the knowledge we acquired about the relative position. Under the conditions considered here the relative position can thus be known and measured to any desired accuracy at any time. Since a relative position is all we can ask for we conclude that it is possible to have quantum trajectories without quantum uncertainties.

Consider now a repeated or time-continuous measurement of the relative position $X-X_0$. If the backaction noise is cancelled the strength of the meter can be increased indefinitely and hence the meter noise contribution can be negligible. In this case the relative coordinate and momentum for the two oscillators can be measured arbitrarily well. The precision of this measurement is given by the entanglement condition $\Var(X-X_0)+\Var(P+P_0)<2$. The equality achieved for uncorrelated systems in a pure state corresponds to two units of vacuum noise $\Var(X-X_0)=\Var(P+P_0)=1$. This benchmark exactly corresponds to the optimal performance attainable in conventional, back action non-evading measurement: There the optimal balance between meter and back action noise achieved at the standard quantum limit consists of one unit due to the system uncertainty, half a unit due to the meter uncertainty and another half unit due to the measurement backaction, that is two units of vacuum noise in total \cite{Hammerer2005,Braunstein2000}. In the back-action evading scheme suggested here entanglement is thus necessary and sufficient for measurements beyond the SQL.
\section*{Noiseless trajectory for an oscillator}
The same logic just discussed for free particles applies also to oscillators. The time dynamics of a harmonic oscillator is given by $X(t)=X(0)\cos(\omega t)+P(0)\sin(\omega t) $ and  $P(t)=  P(0)\cos(\omega t)-X(0)\sin(\omega t)$ where we refer again to dimensionless position and momentum variables. Again, exact knowledge of $X(0)$ corresponding to the squeezed state of the oscillator at $t=0$ leads to the trajectory which is very noisy at $t \neq 2\pi/ \omega, 4\pi/ \omega ...$ due to the quantum noise of backaction of the measurement of $X(0)$ imposed on $P(0)$. As for a free particle, an entangled EPR state between the oscillator of interest and a reference oscillator leads to a noiseless trajectory provided that the reference oscillator has an effectively negative mass. Indeed, in this case  $X(t)-X_0(t)=[X(0)-X_0(0)]\cos(\omega t)+[P(0)+P_0(0)]\sin (\omega t)$. The commuting operators $X(0)-X_0(0)$ and $P(0)+P_0(0)$ can be known exactly and hence $X(t)-X_0(t)$ can be also known exactly at any given time.


\begin{figure}
\center\includegraphics[width=0.7\columnwidth]{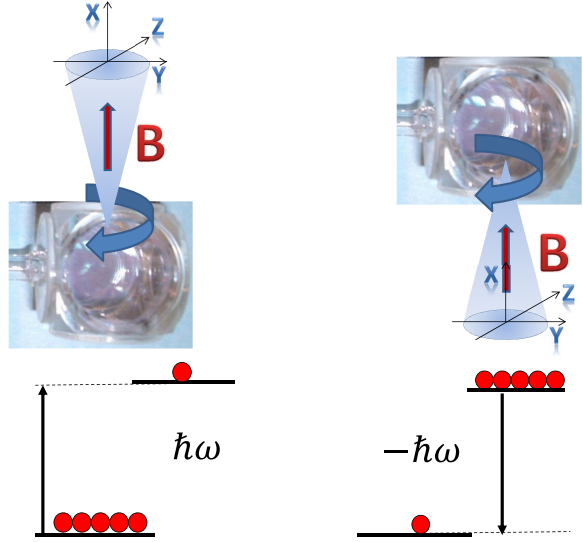}\\
  \caption{Collective spin oscillators with an effective positive and negative masses. The ensemble on the left is polarized along the magnetic field direction. Excitation of the ensemble from a fully polarized state corresponds to the precession of the spin in the field with positive magnetic energy. The ensemble on the right is polarized in the opposite direction and its excitation corresponds to the precession with a negative magnetic energy, or an effective negative mass.}\label{figure_levels}
\end{figure}

As a specific example of two entangled oscillators of which one can exhibit a negative mass we provide an illustration in terms of magnetic oscillators realized in atomic ensembles. Consider two collective magnetic moments (spins) of atomic ensembles \cite{RMP} in a magnetic field, see Fig.~(\ref{figure_levels}). One of the oscillators is oriented along the B-field and hence precesses clockwise when seen along its own mean orientation $J_{x1}=J_x$ (upwards in the Fig.) at the Larmor frequency $\omega$, such that $J_{y1}(t)=J_{y1}(0) \cos(\omega t)+J_{z1}(0)\sin (\omega t) $ and  $J_{z1}(t)=J_{z1}(0) \cos(\omega t)-J_{y1}(0)\sin (\omega t) $. For atoms with positive g-factors such rotation corresponds to positive magnetic energy. The other spin is oriented oppositely to the B-field and hence precesses counter-clockwise when seen along its orientation $J_{x2}=-J_x$ (downwards in the Fig.), such that   $J_{y2}(t)=J_{y2}(0) \cos(\omega t)-J_{z2}(0)\sin (\omega t) $ and  $J_{z2}(t)=J_{z2}(0) \cos(\omega t)+J_{y2}(0)\sin (\omega t) $. Such rotation leads to the negative magnetic energy and can be formally obtained from the expressions for the first spin by changing the sign of the frequency $\omega$. Physically both spins precess clockwise when seen along the common direction of the B-field, that is in this common reference frame
$$J_{y1}(t)+J_{y2}(t)=[J_{y1}(0)+J_{y2}(0)] \cos(\omega t)+[J_{z1}(0)+J_{z2}(0)]\sin (\omega t) $$
and
$$J_{z1}(t)+J_{z2}(t)=[J_{z1}(0)+J_{z2}(0)] \cos(\omega t)-[J_{y1}(0)+J_{y2}(0)]\sin (\omega t). $$
From the spin  commutation relations $[J_{z1}(0),J_{y1}(0)]=-[J_{z2}(0),J_{y2}(0)]=J_x$ where we approximate $J_x$, the spin component along the axis of atomic polarization, as a classical variable. We infer that $[J_{z1}(0)+J_{z2}(0),J_{y1}(0)+J_{y2}(0)]=0$. Hence the initial mutual orientation of the two collective spins can be measured and known beyond the standard quantum limit.


The better than the standard quantum limit correlation of the two spins can be cast in the language of the canonical oscillators by introducing the variables: $X_{1}=J_{y1}/\sqrt{J_x}$, $X_{2}=-J_{y2}/\sqrt{J_x}$ and $P_{1}=J_{z1}/\sqrt{J_x}$, $P_{2}=J_{z2}/\sqrt{J_x}$ which follow the canonical commutation relation $[X_{1(2)},P_{1(2)}]=i$ with variances $\Var(X)=\Var(P)=1/2$ in the minimal uncertainty state called a coherent spin state (CSS). From the above equations we obtain  $X(t)-X_0(t)=[X(0)-X_0(0)]\cos(\omega t)+[P(0)+P_0(0)]\sin (\omega t)$. This relation implies that if an entangled state of these two oscillators with $\Var[X(0)-X_0(0)]\rightarrow 0$ and
$\Var[P(0)+P_0(0)]\rightarrow 0$ is created at $t=0$, the relative canonical coordinate of one oscillator in the reference frame of the other one shall have vanishingly small uncertainty at all times  $\Var[X(t)-X_0(t)]\rightarrow 0$.
\section*{Experiment with magnetic oscillators}
\begin{figure}
\center\includegraphics[width=0.7\columnwidth]{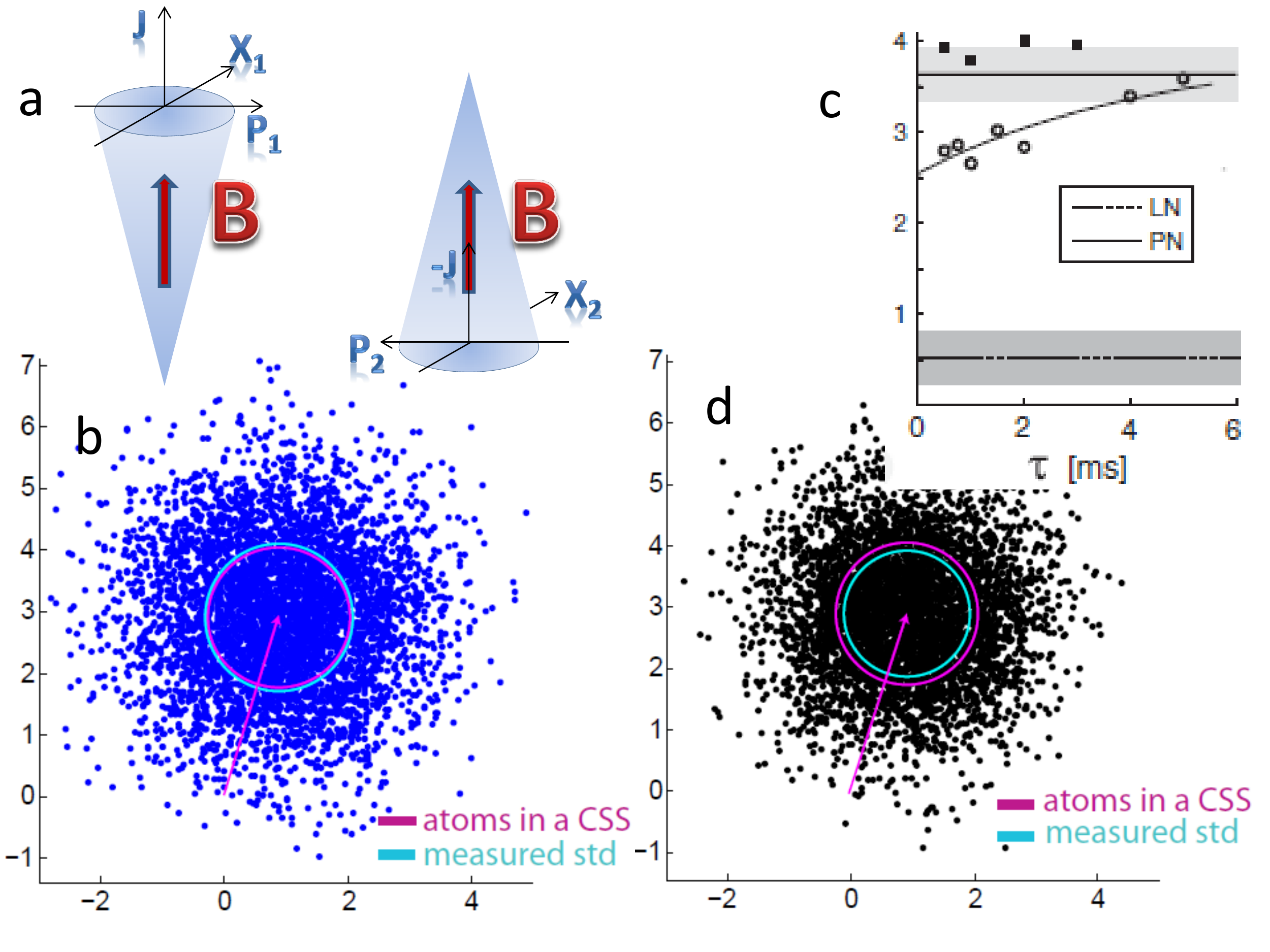}\\
\caption{\textbf{Trajectory beyond the standard quantum limit}. (a) Two magnetic oscillators (b) Results of a series of measurements of 1 msec evolution of one spin in the reference frame of the other spin for both spins for uncorrelated spins (c) The EPR variance of the entangled state of the two spin oscillators (d) Same as in (b) but for an entangled state of the two oscillators }\label{Fig:clouds}
\end{figure}

We can illustrate the effect of entanglement enhanced back action evasion using the data obtained in the experiment \cite{Wasilewski2010}. Consider two magnetic oscillators as introduced above. Initially the two ensembles are prepared in a minimal uncertainty state with opposite macroscopic spin directions as shown in Fig.~(\ref{Fig:clouds}a).  After that a radio-frequency magnetic field is applied to one of the spin oscillators for a time $\tau$, so that it dynamically evolves following a trajectory indicated with an arrow starting from the initial zero point. As shown in \cite{Wasilewski2010}, the coordinate of one of the oscillators with respect to the other one can be measured in a quantum nondemolition way by sending a pulse of light through the two ensembles. The result of the measurement yields the values of $X_{1}-X_{2}=J_{y'}^{1}/\sqrt{J_x}-J_{y'}^{2}/\sqrt{J_x}$ and $P_{1}+P_{2}=J_{z'}^{1}/\sqrt{J_x}-J_{z'}^{2}/\sqrt{J_x}$.  The result of this experiment repeated many times is shown in the left of Figure~\ref{Fig:clouds}b. The red line starting at zero point presents the mean displacement of the oscillator during the time $\tau$. The magenta circle shows the standard deviation of the distribution of the results. It is very close to the purple circle which is the calculated standard deviation for the spins in the minimal uncertainty state. These results demonstrate the measurement of the trajectory of an oscillator with the uncertainty being very close to the standard quantum limit.

In the next experiment an entangled state between the two oscillators is created.  Fig.~\ref{Fig:clouds}c from  \cite{Wasilewski2010} presents measurements of the EPR variance as a function of time after the state is generated. The horizontal solid line corresponds to the projection noise of the spin oscillator. The circles correspond to the entangled state created at $t=0$ by a measurement on the pulse of light propagating through both ensembles. This prepares the necessary EPR entangled state between the positive and negative mass oscillator. The state has the EPR variance $\Delta_{EPR}/2=0.7$ immediately after the state is generated.  After that the radio-frequency magnetic field is applied to one of the spin oscillators for a time $\tau$, so that it has the same classical dynamics as in the first experiment. After that the measurement of $X_{1}-X_{2}$ and $P_{1}+P_{2}$ is performed. Figure~\ref{Fig:clouds}d shows the results of this series of trajectory measurements. The standard deviation of the trajectories is 0.84 of the standard deviation defined by the standard quantum limit for the spin, a clear demonstration of back action evading magnetometry catalyzed by entanglement.

As proposed in ~\cite{Hammerer2009} entanglement can be generated between an atomic spin oscillator with a negative mass described above and a mechanical oscillator coupled to light via radiation pressure. With such an entangled state a trajectory of the mechanical oscillator can be measured with the precision beyond the SQL as has been demonstrated for two spin oscillators.

With measurements of fields, time, acceleration and position approaching or reaching the quantum limits of precision strategies to achieve back action evasion gain practical relevance. The  considerations laid out above show that it is in principle possible to attain  trajectories without quantum uncertainties and measurements of motion beyond standard quantum limits are possible. It is likely that there is another limit to the precision of the measurements described in this paper which is analogous to the Heisenberg limit and hence scales with the number of particles as $1/N$. This limit which is far beyond the present technologies for macroscopic objects discussed here will be analyzed elsewhere.

\emph{Acknowledgements.} We thank K. Jensen for making the Fig. 3. We acknowledge funding by the ERC grant INTERFACE and EU projects SIQS and iQUOEMS.

%

\bibliographystyle{adp}
\bibliography{manuscript_polzik}

\end{document}